\documentclass[reprint,aps,graphicx,prx,showpacs]{revtex4-1}
\usepackage{amsmath,amssymb,amsfonts,mathtools}     

\usepackage{graphicx,bm,MnSymbol}
  \usepackage{paralist}
  \usepackage{epstopdf}
  \usepackage{graphics} 
 \usepackage[colorlinks=true]{hyperref}
 \hypersetup{urlcolor=blue, citecolor=red}
 \usepackage[latin1]{inputenc}
\usepackage[caption=false]{subfig}
\usepackage{soul}

\renewcommand{\theequation}{\arabic{section}.\arabic{equation}}

\newcommand{\e}{{\rm e}}

\renewcommand{\d}{{\rm d}}
\renewcommand{\P}{\mathbb{P}}
\newcommand{\R}{\mathbb{R}}

\newcommand{\E}{\mathbb{E}}

\renewcommand{\L}{{\mathbb L}}

\def\P{\mathbb{P}}

\begin{document}

\title{Trapping of a run-and-tumble particle in an inhomogeneous domain: the weak noise limit}

\author{Paul C. Bressloff}
\address{Department of Mathematics, University of Utah, Salt Lake City, UT 84112 USA}

\begin{abstract}
A one-dimensional run-and-tumble particle (RTP) switches randomly between a left and right moving state of constant speed $v$. This type of motion arises in a wide range of applications in cell biology, including the unbiased growth and shrinkage of microtubules or cytonemes, the bidirectional motion of
molecular motors, and the ``run-and-tumble'' motion of bacteria such as {\em E. coli}. RTPs are also of more general interest within the non-equilibrium statistical physics community, both at the single particle level and at the interacting population level, where it provides a simple example of active matter. In this paper we use asymptotic methods to calculate the mean first passage time (MFPT) for a one-dimensional RTP to escape an effective trapping potential generated by space-dependent switching rates. Such methods are part of a more general framework for studying metastability in so-called piecewise deterministic Markov processes (PDMPs), which include the RTP as a special case.
 
 \end{abstract}

\maketitle

\section{Introduction}

Velocity jump processes, whereby a particle randomly switches between different velocity states, are finding a growing number of applications in cell biology. The particle could represent a bacterial cell such as {\em E. coli} undergoing chemotaxis \cite{Berg77,Berg04,Hillen00}, a motor-cargo complex walking along a cytoskeletal filament \cite{Reed90,Friedman05,Newby10,Newby10a,Newby11,Bressloff13}, the tip of a microtubule undergoing alternating periods of growth and shrinkage (catastrophes) \cite{Dogterom93}, or the tip of a cytoneme filament searching for a target cell during morphogenesis \cite{Bressloff19}. One of the simplest examples of a velocity jump process, is the so called one-dimensional run-and-tumble particle (RTP), which switches between two velocity states $\pm v$. (Within the context of {\em E. coli}, a run refers to a period of almost constant ballistic motion, whereas tumbling is the disordered local motion that selects a new random direction for the next run.)  The run-and-tumble model has also attracted considerable recent attention within the non-equilibrium statistical physics community, both at the single particle level and at the interacting population level, where it provides a simple example of active matter \cite{Tailleur08,Cates15,Volpe16}. Studies at the single particle level include properties of the position density of a free RTP in one and higher dimensions \cite{Martens12,Gradenigo19,Singh19,Santra20a}, first-passage time (FPT) properties \cite{Angelani14,Angelani15,Malakar18,Demaerel18,Scacchi18,Mori19,Doussal19}, RTPs under stochastic resetting \cite{Evans18,Bressloff20a,Santra20}, and non-Boltzmann stationary states for an RTP in a confining potential \cite{Dhar19,Sevilla19,Dor19,Basu20,Doussal20}.

In the case of bacterial run-and-tumble, a chemotactic concentration gradient can bias the tumbling rate so that the bacterium executes motion towards a source of chemoattractant or away from a source of chemorepellant. In order to model one-dimensional chemotaxis using the simplified RTP model, it is necessary to introduce some bias in the stochastic switching (tumbling) between 
the velocity states $\pm v$ that depends 
on the extracellular concentration gradient $c$ \cite{Erban05a,Bialek12}. An alternative RTP modeling paradigm is to assume that switching favors the negative velocity state for $x\rightarrow \infty$ and the positive velocity state for $x\rightarrow -\infty$. The spatially-dependent switching thus acts as an effective confining potential, which can lead to a non-Boltzmann-like stationary probability distribution \cite{Singh21}. 

The role of spatially dependent switching rates has also been explored in a variety of intracellular transport models, including both diffusive transport \cite{Bressloff17,Bressloff19a} and active transport. An example of the latter arises in the so-called tug-of-war model of motor-driven bidirectional transport along microtubules \cite{Gross04,Muller08}. Microtubules are polarized polymeric filaments with biophysically distinct ($+$) and $(-)$ ends, and this polarity determines the preferred direction in which an individual molecular motor moves. In particular, kinesin motors move towards the $(+)$ end whereas dynein motors move towards the $(-)$ end. If both kinesin and dynein motors are attached to a vesicular cargo, then the velocity state will be determined by how many of the kinesin and dynein motors are bound to the microtubule at any one time. In addition, the switching between different velocity states will depend on the rates of binding and unbinding of individual motors to the filament track. 
One mechanism for generating space-dependent transition rates involves microtubule associated proteins (MAPs). These molecules bind to microtubules and effectively modify the free energy landscape of motor-microtubule interactions.  For example, tau is a MAP found in the axon of neurons and is known to be a key player in Alzheimer's disease. Experiments have shown that tau significantly alters the dynamics of kinesin; specifically, by reducing the rate at which kinesin binds to the microtubule \cite{Vershinin07}. This can be interpreted as an effective space-dependent increase in the rate of switching to negative velocity states.

The effect of local tau signaling on a tug-of-war model has been explored in terms of a multi-state velocity jump process with space-dependent switching rates \cite{Newby10a}. Analogous to the more recent study of an RTP \cite{Singh21}, a local increase in the tau concentration acts as an effective confining potential for the motor complex. This can be understood heuristically as follows. When a kinesin driven cargo encounters 
the MAP-coated trapping region the motors unbind at their usual rate and can't rebind. Once the dynein motors are strong enough to pull the remaining kinesin motors off the microtubule, the motor-complex quickly transitions to $(-)$ end directed transport. After the dynein-driven cargo leaves the MAP-coated region, kinesin motors can then re-establish $(+)$ end directed transport until the motor-complex 
returns to the MAP-coated region. This back-and-forth motion repeats until eventually the motor-complex is able to move forward past the MAP-coated region.  
Interestingly, particle tracking experiments have observed oscillatory behavior of motor-driven mRNA particles around synaptic targets in the dendrites of neurons \cite{rook00,dynes07}. This has led to the hypothesis that local tau signaling enhances the probability of a motor complex delivering its vesicular cargo to a target \cite{Newby10a}. The amount of time that the motor complex spends within the target domain can then be formulated as a mean FPT (MFPT) problem \cite{Newby11}. 

One of the assumptions in Ref. \cite{Newby11} is that the switching rates are fast relative to other dynamical processes (weak noise assumption). This means that the escape from the effective confining potential involves rare events that cannot be accurately captured using a diffusion approximation of the velocity jump process. Instead, a combination of the Wentzel-Kramers-Brillouin (WKB) method and matched asymptotics are used to calculate the MFPT. Such methods have also been applied to a more general class of stochastic processes known as piecewise deterministic Markov processes (PDMPs), with particular applications to stochastic ion channels \cite{Keener11,NBK13,Bressloff14b,Newby14}, gene networks \cite{Newby12,Newby13a,Newby15} and stochastic neural networks \cite{Bressloff13a,Bressloff14,Yang19}. A PDMP involves the coupling between a discrete Markov chain $N(t)\in \{0,1,\ldots ,M\}$ and a continuous process $x(t)\in \R^d$ that evolves deterministically between jumps in the discrete random variables \cite{Davis84}. That is, $\dot{x}=F_n(x)$ when $N(t)=n$, where $\{F_n(x),n=0,1,\ldots M-1\}$ is a set of vector fields. A velocity jump process is a special class of PDMP for which $F_n(x)=v_n$, where $v_n$ is the $n$-th velocity state, and a one-dimensional RTP corresponds to the case $M=2$ with $v_0=v,v_1=-v$. 

As far as we are aware, the connection between the statistical physics of RTPs and the more general theory of velocity jump processes and PDMPs has not been explored in any detail.
 In this paper, we show how methods developed to analyze metastability in PDMPs can be used to study a one-dimensional RTP with an effective trapping potential due to space-dependent switching rates. In Sect. II we introduce the basic model and discuss various choices for the transition rates. One of the simplifying features of the model compared to more general PDMPs is that an exact solution for the stationary distribution of the RTP position can be derived without recourse to some approximation scheme such as WKB. The main part of the paper is developed in Sect. III, where we use the asymptotic analysis developed in Ref. \cite{Newby11} to calculate the MFPT for the RTP to escape the effective trapping potential generated by the space-dependent switching rates.

\setcounter{equation}{0}
\section{Run-and-tumble particle with space-dependent switching}

Consider an RTP that randomly switches between two constant velocity states labeled by $n=0,1$ with $v_0=v$ and $v_1=-v$ for some $v>0$. The position $X(t)$ of the particle at time $t$ evolves according to the velocity jump process
\begin{equation}
\label{PDMP}
\frac{dX}{dt}=v[1-2n(t)],
 \end{equation}
where $n(t)=0,1$. Furthermore, suppose that the particle reverses direction according to a two-state Markov chain with space-dependent transition rates
\begin{equation}
0 \xrightleftharpoons[\alpha(x)]{\beta(x)} 1.
\end{equation}
Let $p_{n}(x,t)$ be the probability density of the RTP at position $x\in \R$ at time $t>0$ and moving to the right ($n=0)$ and to the left ($n=1$), respectively. The associated differential Chapman-Kolomogorov (CK) equation is then
\begin{subequations}
\label{DL}
\begin{align}
\frac{\partial p_{0}}{\partial t}&=-v \frac{\partial p_{0}}{\partial x}- \beta(x)  p_{0}+\alpha(x) p_{1},\\
\frac{\partial p_{1}}{\partial t}&=v \frac{\partial p_{1}}{\partial x}+\beta(x) p_{0}-\alpha(x) p_{1}.
\end{align}
\end{subequations}
This is supplemented by the initial conditions $x(0)=x_0$ and $n(0)=n_0$ with probability $\rho_{0,n_0}$ such that $\rho_{0,0}+\rho_{0,1}=1$. In matrix form, we can write
\begin{equation}
\label{DL0}
\frac{\partial p_{n}}{\partial t}=-v_n\frac{\partial p_{n}}{\partial x}+\sum_{n=0,1} Q_{nm}(x)p_m,
\end{equation}
with $v_n=v(1-2n)$ and
\begin{equation}
\label{Q}
{\bf Q}=\left (\begin{array}{cc} -\beta(x) &\alpha(x) \\ \beta(x)& -\alpha(x) \end{array} \right ).
\end{equation}
The matrix version is easily generalizable to more than two velocity states.

Let $L$ be some characteristic distance, which could be related to the space constant of a chemical concentration gradient in chemotaxis, or the size of a target in motor-driven cargo transport. This introduces a natural time scale $T=L/v$. Suppose that the transition rates $\alpha(x),\beta(x) \gg 1/T$ for all $x\in \R$, so that we can we can take $\alpha,\beta=O(1/\epsilon)$ on relevant length and time scales. Rescaling the transition rates in Eq. (\ref{DL0}) thus gives
\begin{align}
\label{DL2}
\frac{\partial p_{n}}{\partial t}&=-v_n \frac{\partial p_{n}}{\partial x}+\frac{1}{\epsilon} \sum_{m=0,1}Q_{nm}(x)p_m.
\end{align}
For a given $x$, define the average velocity
\begin{equation}
\label{avev}
V(x)=v\rho_0(x)-v\rho_1(x),
\end{equation}
where
\begin{equation}
\rho_0(x)=\frac{\alpha(x)}{\alpha(x)+\beta(x)},\quad \rho_1(x)=1-\rho_0(x)
\end{equation}
is the stationary probability distribution of the two-state Markov chain with generator ${\bf Q}(x)$, that is, $\sum_{m=0,1}Q_{nm}(x)\rho_m(x)=0$. Intuitively speaking, one expects Eq. (\ref{PDMP}) to reduce to the deterministic dynamical system 
\begin{equation}
\label{mft}
\frac{dx(t)}{dt} =  V(x(t)),\quad
x(0) = x_0
\end{equation}
 in the fast switching or adiabatic limit $\varepsilon \rightarrow 0$. That is, for sufficiently small $\varepsilon$, the Markov chain undergoes many jumps over a small time interval $
\Delta t$ during which $\Delta x\approx 0$, 
and thus the relative frequency of each discrete state $m$ is approximately $\rho_m(x)$. This can be made precise in terms of a law of large numbers for velocity jump processes, as well as more general PDMPs \cite{Kifer09,fagg09,Faggionato10,Pakdaman12}.

\subsection{Concentration gradient} One possible source of space-dependent switching or tumbling rates is a chemical concentration gradient $c(x)$. For the sake of illustration, consider a simple phenomenological model, in which the tumbling rates depend on the time derivative of the 
concentration $c(t)=c(x(t))$ along the 
particle trajectory, where $x(t)$ is the particle position at time $t$ \cite{Bialek12}. Using the fact that $\dot{c}=\pm v dc/dx$, we take
\begin{subequations}
\label{RT}
\begin{align}
\beta(x) &=k_0+ k_1vc'(x),\quad \alpha(x)= k_0- k_1vc'(x).
\end{align}
\end{subequations}
(For simplicity, switching depends on the instantaneous value of the concentration gradient rather than a time averaged change in 
concentration as is typical in bacterial chemotaxis \cite{Berg77}.)
 The stationary probability densities satisfy the pair of equations
 \begin{align*}
 v\frac{d p_0}{d x}&=- \beta(x) p_0(x)+\alpha(x) p_1(x),\\
 -v\frac{\d p_1}{dx}&= \beta(x) p_0(x)-\alpha(x) p_1(x).
 \end{align*}
Adding these two equations gives
\[v\frac{dp_0}{d x}-v\frac{d p_1}{d x}=0,\]
which implies that the difference $p_0(x)-p_1(x)=\mbox{constant}$. Assuming that $-\infty<x<\infty$, normalizability of the probability densities requires this constant to 
be zero. Hence, $p_{0,1}(x)=p(x)/2$ with 
$p(x)$ satisfying the single equation
\[v\frac{d p}{d x}=\left [\alpha(x) -\beta(x) \right ]p(x)=-2k_1vc'(x)p(x).\]
This has the straightforward solution
\begin{equation}
\label{chempp}
 p(x)={\mathcal N}\e^{-2k_1c(x)},
\end{equation}
where ${\mathcal N}$ is a normalization factor. If the signaling molecules correspond to a chemoattractant then the rate of tumbling decreases in the direction for which $\dot{c}
>0$, that is, $k_1<0$, and maxima of the 
stationary solution (\ref{chempp}) coincide with maxima of the concentration $c(x)$. Conversely, $k_1>0$ for a chemorepellant and maxima of $p(x)$ coincide with 
minima of the concentration.

\subsection{Localized trap}

\begin{figure}[b!]
\centering
\includegraphics[width=7cm]{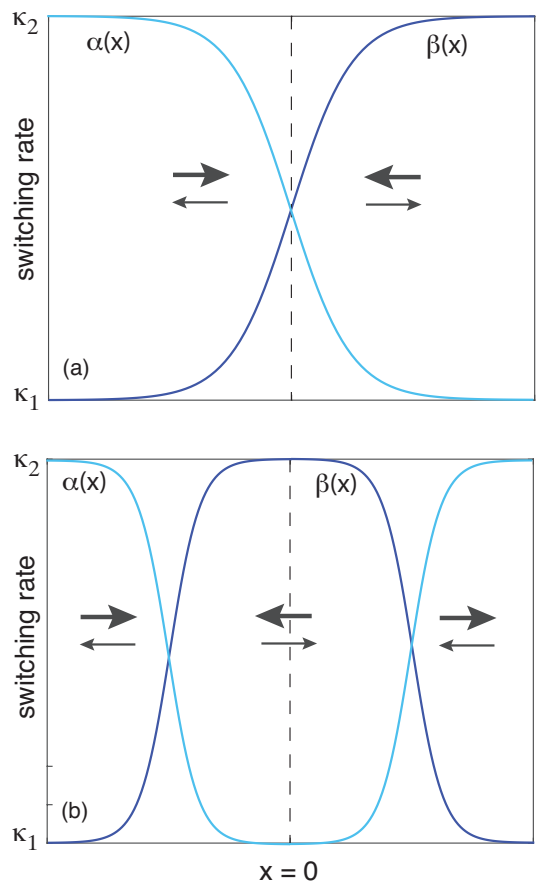}
\caption{Space-dependent switching rates $\alpha(x),\beta(x)$ for a one-dimensional RTP. (a) Switching rates (\ref{trap}). (b) Switching rates (\ref{trap1}). Thickness of the arrows indicates preferred velocity direction.}
\label{fig1}
\end{figure}

In this paper we are interested in a different form of space-dependent switching, namely one that traps the RTP within a local region which, without loss of generality, we take to be in a neighborhood of the origin. We will consider two different examples of trapping mechanisms as illustrated in Fig. 1. The first  mechanism favors the right-moving velocity when $x<0$ and the left-moving velocity state when $x>0$, Fig. \ref{fig1}(a). This can be implemented using the switching rates
\begin{subequations}
\label{trap}
\begin{align}
 \alpha(x)&=\kappa_2+\frac{1}{2}(\kappa_1-\kappa_2)(1+\tanh (x/\gamma)),\\ \beta(x)&=\kappa_1+\frac{1}{2}(\kappa_2-\kappa_1)(1+\tanh (x/\gamma)).
\end{align}
\end{subequations}
Clearly $(\alpha(x),\beta(x))\rightarrow (\kappa_1,\kappa_2)$ as $x\rightarrow  \infty$ and $(\alpha(x),\beta(x))\rightarrow (\kappa_2,\kappa_1)$ as $x\rightarrow  -\infty$. In addition, the sharpness of the transition is determined by $\gamma$ such that in the limit $\gamma \rightarrow 0$, 
\begin{subequations}
\label{trapI}
\begin{align}
\beta(x)&=\kappa_1+(\kappa_2-\kappa_1)\Theta(x),\\ \alpha(x)&=\kappa_2+ (\kappa_1-\kappa_2)\Theta(x),
\end{align}
\end{subequations}
where $\Theta(x)$ is the Heaviside function. The average velocity (\ref{avev}) is given by
\begin{equation}
V(x)=\frac{\kappa_1-\kappa_2}{\kappa_1+\kappa_2} \tanh(x/\gamma)v.
\end{equation}
Hence, if $\kappa_1<\kappa_2$ then $x=0$ is a globally attracting fixed point of the deterministic system (\ref{mft}),
suggesting that the RTP tends to be localized around the origin, at least in the weak noise regime. Note that Eq. (\ref{trapI}) is one of the few  space-dependent transition rates for which an exact solution of the time-dependent probability density $p(x,t)$ can be derived without restricting to the weak noise regime \cite{Singh21}.

The second mechanism assumes that the left-moving state is favored in a local region of the origin, whereas the right-moving state is favored on either side of this domain, Fig. \ref{fig1}(b). The corresponding switching rates are taken to be of the from
\begin{subequations}
\label{trap1}
\begin{align}
 \alpha(x)&=\kappa_1+(\kappa_2-\kappa_1)\tanh^2 (x/\gamma)),\\ \beta(x)&=\kappa_1+(\kappa_2-\kappa_1)(1-\tanh^2 (x/\gamma)).
\end{align}
\end{subequations}
It can be seen that $\alpha(0)=\kappa_1,\beta(0)=\kappa_2$, whereas $\alpha(x)\rightarrow \kappa_2,\beta(x)\rightarrow \kappa_1$ as $|x|\rightarrow \infty$.
Moreover, the average velocity (\ref{avev}) is 
\begin{equation}
V(x)=v\frac{\kappa_2-\kappa_1}{\kappa_1+\kappa_2} (1-2\mbox{sech}^2(x/\gamma)).
\end{equation}
Now there are two fixed points at $x=\pm \bar{x}$ where
\begin{equation}
\label{barx}
\tanh(\bar{x}/\gamma)=\frac{1}{\sqrt{2}}.
\end{equation}
If $\kappa_1<\kappa_2$, then the fixed point $-\bar{x}$ is stable and the fixed point $\bar{x}$ is unstable. (The second mechanism is analogous to the trapping of a molecular motor complex by a local region of enhanced tau concentration \cite{Newby10a}, which was described in the introduction.)

\begin{figure}[t!]
\centering
\includegraphics[width=8cm]{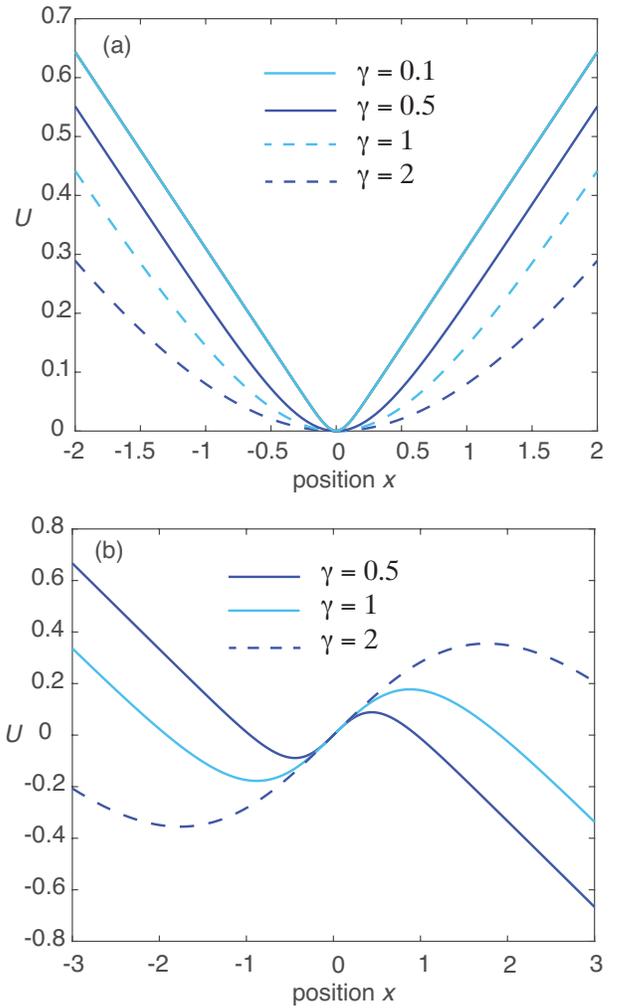}
\caption{Deterministic potential $U(x)$ corresponding to (a) the switching rates (\ref{trap}) and (b) the switching rates (\ref{trap1}) for various gains $\gamma$. Other parameters are $v=1$, $\kappa_1=0.5$ and $\kappa_2=1$.}
\label{fig2}
\end{figure}

The differences between the two cases becomes clearer by
rewriting the deterministic Eq. (\ref{mft}) as the gradient system
\begin{equation}
\frac{dx}{dt}=-\frac{dU(x)}{dx}, 
\end{equation}
with
\begin{equation}
\label{U}
U(x)=v\gamma \frac{\kappa_2-\kappa_1}{\kappa_1+\kappa_2}\ln \cosh(x/\gamma),
\end{equation}
for the switching rates (\ref{trap}) and
\begin{equation}
\label{U1}
U(x)=v\frac{\kappa_2-\kappa_1}{\kappa_1+\kappa_2}[-x+2\gamma \tanh(x/\gamma)]
\end{equation}
for the switching rates (\ref{trap1}). As shown in Fig. \ref{fig2}(a), Eq. (\ref{U}) corresponds to a global, symmetric potential well that has a unique minimum at $x=0$, and sharpens as $\gamma \rightarrow 0$. On the other hand, the potential of Eq. (\ref{U1}) is a cubic that is characterized by a potential well in the domain $(-\infty,\bar{x})$ with a minimum at $-\bar{x}$ and barrier height $U(\bar{x})-U(-\bar{x})$, see Fig. \ref{fig2}(b). The deterministic potential $U(x)$ also determines the steady-state solution of the CK Eq. (\ref{DL2}). That is, the steady-state solution is $p_0(x)=p_1(x)=p(x)/2$ with
\begin{equation}
\label{sswkb}
p(x)={\mathcal N}\e^{-\Phi(x)/\epsilon},
\end{equation}
where ${\mathcal N}$ is a normalization factor and
\begin{align}
\Phi(x)&=-\int_0^x\frac{\alpha(y)-\beta(y)}{v} dy\nonumber \\
&=-\frac{\kappa_1+\kappa_2}{v^2}\int_0^xV(y)dy=\frac{\kappa_1+\kappa_2}{v^2}U(x).
\label{QPwkb}
\end{align}
One can identify $\Phi(x)$ as the so-called quasipotential.

\begin{figure}[b!]
\centering
\includegraphics[width=7cm]{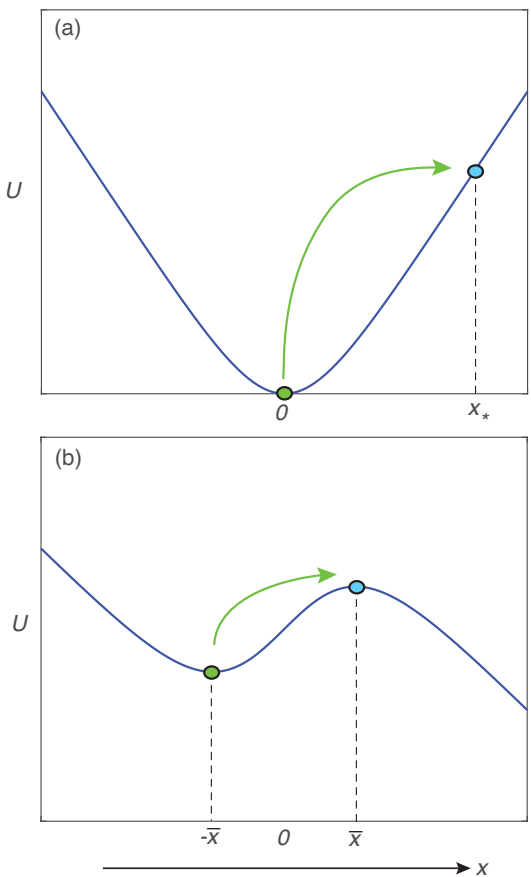}
\caption{FPT problems for the (a) unimodal potential and (b) cubic potential.}
\label{fig3}
\end{figure}

\subsection{First passage time (FTP) problem}

Given the potentials $U(x)$ of Fig. \ref{fig2}, we would like to determine the mean time for the RTP to escape a neighborhood of the origin in the weak noise regime. In the case of the unimodal potential (\ref{U}), we consider the MFPT for the particle to reach a location $x_*\gg 0$ far to the right of the origin, given that it started at $x=0$. On the other hand, for the cubic potential (\ref{U1}), we consider the MFPT to reach the maximum $\bar{x}$ given that the particle started at $-\bar{x}$. These two escape problems are illustrated in Fig. \ref{fig3}. In both cases we use asymptotic methods developed for general velocity jump processes \cite{Newby11}. In particular, we show that considerable simplification occurs in the case of an RTP where, for example, the exact stationary density (\ref{sswkb}) is known without any recourse to approximation schemes such as WKB. (The two-state velocity jump process was not explicitly considered in \cite{Newby11}.) Moreover, we highlight a subtle feature of the asymptotic analysis of the cubic potential, arising from the fact that the escape point is a maximum of the potential, see also \cite{Newby13a}. Note that one constraint on the use of asymptotic methods is that the quasipotential is twice differentiable. Hence, space-dependent switching rates such as Eq. (\ref{trapI}) would need to be regularized by replacing the Heaviside function with a sharp sigmoid function.

\setcounter{equation}{0}
\section{Asymptotic analysis of the MFPT}

Consider the RTP with switching rates given by Eq. (\ref{trap}) or (\ref{trap1}). In order to calculate the MFPT to escape a neighborhood of the origin, we  supplement the CK equation (\ref{DL2}) by the absorbing boundary condition 
\begin{equation}
p_1(x_*,t)=0.
\end{equation}
Note that the absorbing boundary condition is only imposed on the component $p_1$ associated with the negative velocity, ensuring that once the RTP reaches $x_*$ it can never reenter the domain. In the case of the unimodal potential we take $x_*\gg 0$, whereas for the cubic potential we set $x_*=\bar{x}$, see Fig. \ref{fig3}. Let $T$ denote the (stochastic) FPT for which the system first reaches $x_*$, given that it started at $x=x_0$. For the unimodal case $x_0=0$, whereas for the cubic case $x_0=-\bar{x}$. The distribution of FPTs is related to the survival probability that the system hasn't yet reached $x_*$, that is, 
\begin{equation*}
\P\{t>T\}=S(t)\equiv \int_{-\infty}^{x_*} \sum_{n=0,1} p_n(x,t)dx .
\end{equation*}
The FPT density is then
\begin{equation}
f(t)=-\frac{dS}{dt}=-\int_{-\infty}^{x_*}\sum_{n= 0,1} \frac{\partial p_n(x,t)}{\partial t}dx .
\label{ogg}
\end{equation}
Substituting for $\partial p_n/\partial t$ using the CK equation (\ref{DL2}) and noting that $\sum_{n}{Q}_{nm}(x)=0$, shows that
\begin{eqnarray}
f(t)&=&\int_{-\infty}^{x_*} \left [\sum_{n=0,1}v_n\frac{\partial p_n(x,t)}{\partial x}\right ]dx\nonumber \\
&=& \sum_{n =0,1}v_np_n(x_*,t)=vp_0(x_*,t)\equiv J(x_*,t),
\label{fTP}
\end{eqnarray}
where $J(x_*,t)$ is the probability flux through the absorbing boundary. 

\subsection{Quasistationary approximation}

Consider an eigenfunction expansion of the time-dependent solution,
\begin{equation}
{\bf p}(x,t)=\sum_{j=0}^{\infty} C_j(t){\bm \phi}_j(x),
\end{equation}
where the eigenfunction ${\bm \phi}_j=(\phi_{j,0},\phi_{j,1})^{\top}$ satisfies the matrix operator equation
\begin{equation}
\L {\bm \phi}_j\equiv \mbox{diag}(v,-v) \frac{\partial {\bm \phi}_j(x)}{\partial x} -\frac{1}{\epsilon}{\bf Q}(x){\bm \phi}_j=\lambda_j {\bm \phi}_j,
\end{equation}
together with the boundary condition 
\begin{equation}
\phi_{j,0}(x_*)=0.
\end{equation}
Here $\mbox{diag}(a,b)$ denotes the diagonal matrix with eigenvalues $a,b$, 
Similarly, we define a set of eigenfunctions for the adjoint operator $\L^{\dagger}$ given by
\begin{equation}
\label{adj}
\L^{\dagger} {\bm \xi}_j\equiv -\mbox{diag}(v,-v) \frac{\partial {\bm \xi}_j(x)}{\partial x} -\frac{1}{\epsilon}{\bf Q}^{\top}(x){\bm \xi}_j=\lambda_j {\bm \xi}_j,
\end{equation}
and the boundary condition
\begin{equation}
\xi_{j,1}(x_*)=0.
\end{equation}
The two sets of eigenfunctions form a biorthogonal set according to the inner product rule
\begin{equation}
\langle {\bm \xi}_j,{\bm \phi}_k\rangle \equiv \int_{-\infty}^{x_*} \sum_{n=0,1} \xi_{j,n}(x)\phi_{k,n}(x)dx=\delta_{j,k}.
\end{equation}
Substituting the eigenvalue expansion into Eq. (\ref{DL2}) shows that the coefficients evolve according to the decoupled equations
\begin{equation}
\frac{dC_j(t)}{dt}=-\lambda_j C_j(t).
\end{equation}

If the absorbing boundary at $x_*$ is replaced by a reflecting boundary, then there is a single zero eigenvalue whose corresponding eigenfunction is the stationary solution on the domain $(-\infty,x_*)$, and all other eigenvalues have positive real parts. We can thus introduce the ordering
\[0=\lambda_0<\mbox{Re}[\lambda_1]\leq \mbox{Re}[\lambda_2]\leq \ldots
\]
On the other hand, when there is an absorbing boundary, the stationary solution no longer exists due to an exponentially small probability flux leaving the system at $x_*$. (This assumes that $0 < \epsilon \ll 1$ so escape is dominated by rare events.) It follows that $\lambda_0$ is perturbed from zero, becoming an exponentially small positive principal eigenvalue:
\[0<\lambda_0\ll \mbox{Re}[\lambda_1]\leq \mbox{Re}[\lambda_2]\leq \ldots
\]
Hence, on intermediate time scales for which the probability of escape is still negligible, contributions from all eigenvalues $\lambda_j$, $j\geq 1$, have decayed to zero and we can make the quasistationary approximation
\begin{equation}
\label{qa}
p_n(x,t)\sim C_0(t){ \phi}_{0,n}(x),\quad n=0,1 .
\end{equation}
In addition, $\phi_0(x)$ is almost identical to the stationary solution (\ref{sswkb}) outside a neighborhood of $x_*$, so that
\begin{equation}
\phi_{0,0}(x)=\phi_{0,1}(x)= \phi_{\epsilon}(x),\quad \phi_{\epsilon}(x)=\e^{-\Phi(x)/\epsilon},
\end{equation}
with $\Phi(x)$ given by Eq. (\ref{QPwkb}). Clearly, the quasistationary solution breaks down around $x_*$ since it does not satisfy the absorbing boundary condition.

It can be checked that under the quasistationary approximation the solution $C_0(t)=C_0(0)\e^{-\lambda_0t}$ still holds. This follows from taking the inner product of Eq. (\ref{DL2}) with the adjoint eigenvector ${\bm \xi}_0$ and substituting for $p_n(x,t)$ using the quasistationary approximation:
\begin{align}
&\left \langle {\bm \xi}_0,\frac{\partial {\bf p}}{\partial t}\right \rangle=-\langle {\bm \xi}_0,\L{\bf p}\rangle \nonumber \\
&\Rightarrow \dot{C}_0\langle {\bm \xi}_0,{\bm \phi}_{\epsilon}\rangle =-\langle\L^{\dagger} {\bm \xi}_0,{\bm \phi}_{\epsilon}\rangle =
-\lambda_0 \langle {\bm \xi}_0,{\bm \phi}_{\epsilon}\rangle,
\end{align}
that is, $\dot{C}_0=-\lambda_0C_0$.
Hence, substituting the quasistationary solution into Eq. (\ref{ogg}) gives
\begin{align}
f(t) &\sim C_0(0)\lambda_0 \e^{-\lambda_0 t} \int_{-\infty}^{x_*}\phi_{\epsilon}(x)dx.
\label{FT}
\end{align}
The constant $C_0(0)$ can be determined from the initial condition $p_n(x,0)=\delta(x)\delta_{n,n_0}$,
and the projection of the eigenfunction expansion onto the adjoint eigenfunction ${\bf \xi}_0$:
\[ \langle {\bm \xi}_0,C_0(0){\bm \phi}_{\epsilon}(x)\rangle=\langle {\bm \xi}_0,{\bf p}(x,0)\rangle =\xi_{0,n_0}(0).\]
In the case of a reflecting boundary at $x_*$, the adjoint eigenfunction ${\bm \xi}_0=(1,1)$. This will still hold in the bulk of the domain for an absorbing boundary at $x_*$ so that we can take
\begin{equation}
C_0(0)=\left [\int_{-\infty}^{x_*}\phi_{\epsilon}(x)dx\right ]^{-1}.
\end{equation}
This establishes that under the quasistationary approximation
\begin{equation}
f(t)\sim \lambda_0\e^{-\lambda_0 t},
\end{equation}
and $\lambda_0^{-1}$ can be identified as the MFPT to escape at $x=x_*$.

In summary, the calculation of the MFPT reduces to the problem of estimating the principal eigenvalue $\lambda_0$. If the exact eigenfunctions ${\bm \phi}_0$ and ${\bm \xi}_0$ were known then we could use either of the inner product identities
\begin{equation}
\lambda_0\langle {\bm \xi}_0,{\bm \phi}_0\rangle = \langle \L^{\dagger}{\bm \xi}_0,{\bm \phi}_0\rangle,\mbox{ or } \lambda_0\langle {\bm \xi}_0,{\bm \phi}_0\rangle = \langle {\bm \xi}_0,\L {\bm \phi}_0\rangle.
\end{equation}
On the other hand, simultaneously using the approximations ${\bm \xi}_0 =(1,1)$ and ${\bm \phi}_0=
{\bm \phi}_{\epsilon} $ yields $\lambda_0=0$, reflecting the breakdown of the quasistationary approximation at the boundary. Therefore, we only apply the quasistationary approximation to ${\bm \phi}_0$ so that
\begin{align}
\lambda_0&\sim \frac{\langle \L^{\dagger}{\bm \xi}_0,{\bm \phi}_{\epsilon}\rangle}{ \langle {\bm \xi}_0,{\bm \phi}_{\epsilon}\rangle}.
\end{align}
Substituting for $\L^{\dagger}$, using integration by parts on the domain $(-\infty,x_*]$, and using $\L\phi_{\epsilon}=0$, shows that
\begin{align}
\label{lam1}
\lambda_0&\sim -\frac{v{\phi}_{\epsilon}(x_*)[\xi_{0,0}(x_*)-\xi_{0,1}(x_*)]}
{ \langle {\bm \xi}_0,{\bm \phi}_{\epsilon}\rangle}.
\end{align}
Following \cite{Newby11}, the adjoint eigenfunction ${\bm \xi}_0(x)$ can be approximated using singular perturbation methods. It is at this stage that escape from the unimodal and cubic potentials have to be treated separately.

\subsection{Calculation of principal eigenvalue: unimodal potential} In order to construct an approximate solution that also satisfies the absorbing boundary condition, we construct a boundary layer in a neighborhood of $x_*$ by performing the change of variables $x=x_*-\epsilon z$ and setting $A_n(z)={\xi}_{0,n}(x_*-\epsilon z)$. Eq. (\ref{adj}) for $j=0$ then becomes to leading order
\begin{equation}
\label{xi1}
v_n\frac{dA_n(z)}{d z}-\sum_{m=0,1}Q_{mn}(x_*)A_m(z)=0,
\end{equation}
together with the boundary condition
\begin{equation}
A_1(x_*)=0.
\end{equation}
This inner solution has to be matched with the outer solution ${\bm \xi}_0={\bf 1}$, which means that
\begin{equation}
\lim_{z\rightarrow \infty}A_n(z)=1,\quad n=0,1.
\end{equation}
Consider the eigenvalue equation
\begin{equation}
\sum_{m=0,1} S_m Q_{mn}(x)v_n^{-1}=\mu  S_{n}.
\label{eig}
\end{equation}
One solution is ${\bf S}_{0}=(1,1)$ and $\mu_0=0$, whereas the other is ${\bf S}_{1}(x)=(\beta(x),\alpha(x))$ and $ \mu_1=-\Phi'(x)=(\alpha(x)-\beta(x))/v$. We now expand the solution $A_n(z)$ in terms of the pair of eigenfunctions at $x=x_*$:
\begin{equation}
\label{An}
A_n(z)=c_0+c_1{S}_{1,n}(x_*)\e^{-\Phi'(x_*)z}.
\end{equation}
Since $\Phi'(x_*) >0$ for $x_*>0$ in the case of the unimodal potential, Fig. \ref{fig3}(a), we see that $A_n(z)\rightarrow c_0$ as $z\rightarrow \infty$, which implies $c_0=1$. The constant $c_1$ is then determined from the boundary condition $A_1(0)=0$:
\begin{equation}
c_1=-\frac{1}{\alpha(x_*)}.
\end{equation}
It follows that
\begin{equation}
\xi_{0,0}(x_*)-\xi_{0,1}(x_*)=\frac{\alpha(x_*)-\beta(x_*)}{\alpha(x_*)}.
\end{equation}

\begin{figure}[t!]
\centering
\includegraphics[width=8cm]{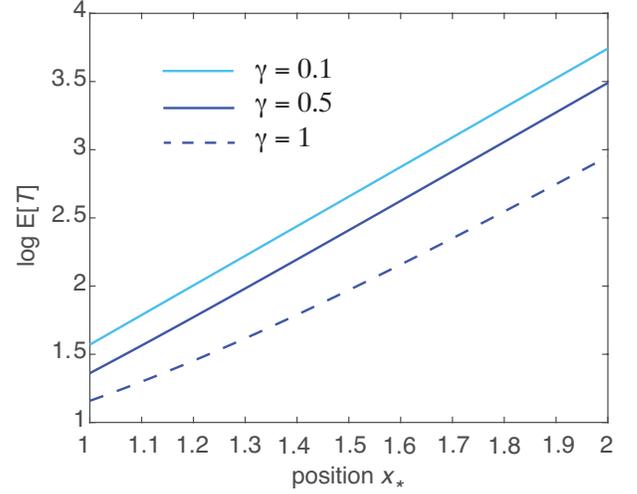}
\caption{Plot of $\log \E[T]$ for the unimodal potential as a function of escape position $x_*$ and various gains $\gamma$. The MFPT $\E[T]=\lambda_0^{-1}$ with $\lambda_0$ given by Eq. (\ref{lam3}). Other parameter values are $\kappa_1=0.5$, $\kappa_2=1$, $v=1$, and $\epsilon=0.1$.}
\label{fig4}
\end{figure}

Substituting the expressions for $\phi_{\epsilon}(x_*)$ and $\xi_{0,0}(x_*)-\xi_{0,1}(x_*)$ into Eq. (\ref{lam1})
and simplifying the denominator using the outer solution ${ \xi}_{0,n} \sim 1$, we obtain the result
\begin{equation}
\lambda_0\sim {\mathcal N}v\frac{\beta(x_*)-\alpha(x_*)}{\alpha(x_*)}\e^{-\Phi(x_*)/\epsilon},
\end{equation}
where
\begin{equation*}
{\mathcal N}=\left [\int_{-\infty}^{x_*}\exp \left (-\frac{\Phi(x)}{\epsilon}\right )\right ]^{-1}.
\end{equation*}
The latter can be approximated using Laplace's method to give
\begin{equation}
\label{norm}
{\mathcal N}\sim \sqrt{\frac{\Phi''(x_0)}{2\pi \epsilon}}\exp \left (\frac{\Phi(x_0)}{\epsilon}\right ).
\end{equation}
Hence, we obtain the following expression for the inverse MFPT:
\begin{equation}
\label{lam3}
\lambda_0\sim  v\frac{\beta(x_*)-\alpha(x_*)}{\alpha(x_*)}\sqrt{\frac{\Phi''(x_0)}{2\pi \epsilon} }\e^{-(\Phi(x_*)-\Phi(x_0))/\epsilon}.
\end{equation}
Setting $x_0=0$ and substituting for the switching rates and the quasipotential using Eqs. (\ref{trap}), (\ref{U}) and (\ref{QPwkb}), we plot $\E[T]=\lambda_0^{-1}$ as a function of the escape position $x_*$ for various degrees of sharpness $\gamma$. The results are shown in Fig. \ref{fig4}.

\subsection{Calculation of principal eigenvalue: cubic potential} 

\begin{figure}[t!]
\centering
\includegraphics[width=8cm]{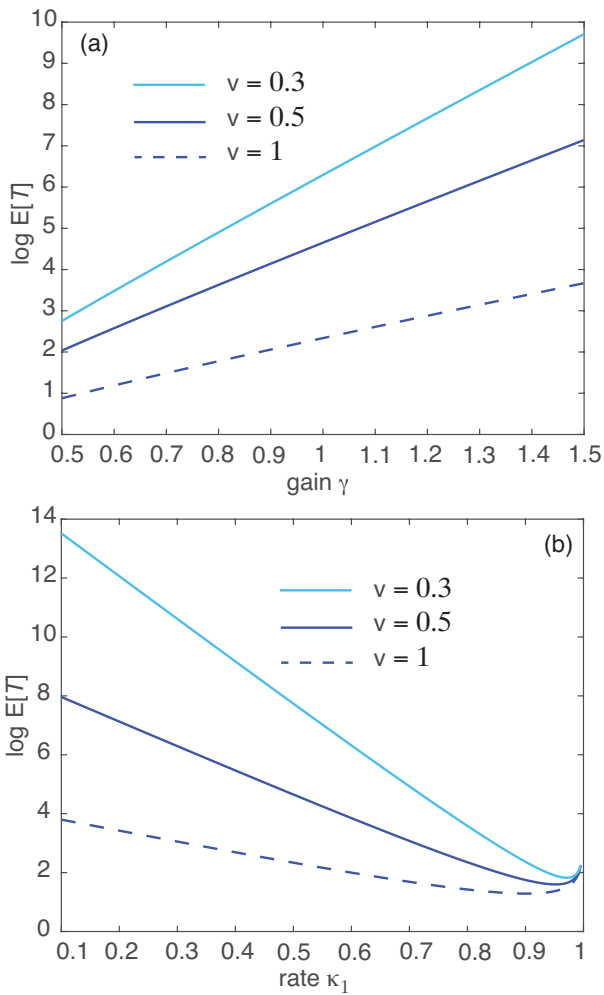}
\caption{Plot of $\log \E[T]$ for the cubic potential as a function of (a) the gain $\gamma$ and (b) the rate $\kappa_1$. The MFPT is $\E[T]=\lambda_0^{-1}$ with $\lambda_0$ given by Eq. (\ref{lam4}). Baseline parameter values are $\gamma=1$, $\kappa_1=0.5$, $\kappa_2=1$, $v=1$, and $\epsilon=0.1$.}
\label{fig5}
\end{figure}

The boundary layer analysis of the unimodal potential breaks down in the case of the cubic potential due to the fact that $\Phi'(x_*)=0$; this is a consequence of the escape point $x_*$ being a local maximum of the potential. In particular, the eigenfunction expansion (\ref{An}) no longer holds since the zero eigenvalue is doubly degenerate at $x=x_*$. Hence, the solution needs to include a secular term involving the generalized eigenvector $\widehat{S}$,
\begin{equation}
\label{ztop}
\sum_{n=0,1}Q_{mn}(x_*)\widehat{S}_m(x_*)=-v_n,
\end{equation}
which implies that
\begin{equation}
\widehat{S}_0(x_*)-\widehat{S}_1(x_*)=\frac{v}{\alpha(x_*)}.
\end{equation}
Note that the Fredholm alternative theorem ensures that $\widehat{S}$ exists and is unique, since the stationary distribution $\rho_m(x_*)$ is the right null vector of ${\bf Q}(x_*)$ and $\sum_{n=0,1}\rho_n(x_*)v_n \equiv V(x_*)=0$; the latter reflects the fact that $x_*=\bar{x}$ is a fixed point of the deterministic equation (\ref{mft}). The solution for ${\bf Q}(z)$ is now
\begin{equation}
\label{An1}
A_n(z)=c_0 +c_1(\widehat{S}_n(x_*)-z).
\end{equation}
The presence of the secular term means that the solution is unbounded in the limit $z\rightarrow \infty$, which implies that the inner solution cannot be matched with the outer solution. One way to remedy this situation is to introduce an alternative scaling in the boundary layer of the form $x=x_*+\epsilon^{1/2}z$,  as detailed in Ref. \cite{Newby12}. One can then eliminate the secular term $-c_1z$ and show that, see appendix A,
\begin{align}
c_0=1-\bar{c_1}\sqrt{\frac{\pi}{2|\Phi''(x_*)|}},\quad c_1=\sqrt{\epsilon}\bar{c}_1,
\end{align}
with $\bar{c}_1$ determined by imposing the boundary condition $A_1(0)=0$:
\begin{equation}
\label{c1}
\bar{c}_1 \sim -\sqrt{\frac{2|\Phi''(x_*)|}{\pi}}+{\mathcal O}(\epsilon^{1/2}),
\end{equation}
Substituting the expressions for $\phi_{\epsilon}(x_*)$ and $\xi_{0,0}(x_*)-\xi_{0,1}(x_*)$ into Eq. (\ref{lam1}), simplifying the denominator using the outer solution ${ \xi}_{0,n} \sim 1$ and Eq. (\ref{norm}), we obtain the result
\begin{equation}
\label{lam4}
\lambda_0\sim \frac{1}{\pi}\frac{v}{\alpha(\bar{x})}\sqrt{ \Phi''(-\bar{x})|\Phi''(\bar{x})|}\e^{-[\Phi(\bar{x})-\Phi(-\bar{x})/\epsilon},
\end{equation}
We have also set $x_*=\bar{x}$ and $x_0=-\bar{x}$ with $\bar{x}$ determined by Eq. (\ref{barx}). Example plots of the MFPT $\E[T]=\lambda_0^{-1}$ are shown in Fig. \ref{fig5} for $\Phi$ given by Eqs. (\ref{U1}) and (\ref{QPwkb}), and $\alpha(x)$ obtained from Eq. (\ref{trap1}). The MFPT is a monotonically increasing function of the gain $\gamma$, since the barrier height increases with $\gamma$:
\begin{equation}
\Phi(\bar{x})-\Phi(-\bar{x})= \frac{2\gamma(\kappa_2-\kappa_1)}{v}(\sqrt{2}-\tanh^{-1}(1/\sqrt{2})).
\end{equation}
Similarly, the MFPT is a decreasing function of the rate $\kappa_1$ as the barrier height becomes smaller as $\kappa_1$ approaches $\kappa_2$. The nonmonotonic behavior of $\E[T]$ for $\kappa_1\approx \kappa_2$ indicates a breakdown of the asymptotic analysis when the barrier height becomes too small.

\setcounter{equation}{0}
\section{Discussion}
In this paper we exploited the connection between RTPs and more general velocity jump processes in order to calculate the MFPT for the RTP to escape from an effective trapping potential in the weak noise limit. In particular, following previous studies of motor-driven bidirectional transport, we showed how the inverse MFPT can be identified with the principal eigenvalue $\lambda_0$ of the CK evolution operator. We then calculated $\lambda_0$ using asymptotic analysis, in order to match the quasistationary solution in the bulk of the domain with an absorbing boundary at the escape point. We also highlighted subtle differences between the unimodal and cubic trapping potentials.

One issue that we did not address is to what extent one can investigate the behavior of the RTP in the weak noise regime using a quasi-steady-state (QSS) or adiabatic approximation. It is well known that in the adiabatic limit, the CK equation of a velocity jump process or a more general PDMP can be approximated by a Fokker-Planck (FP) equation for the total density $p= p_0+p_1$
\cite{Papanicolaou,Reed90,Schnitzer93,Hillen00,Friedman05,Tailleur09,Newby10,Newby10a,Pakdaman12}. The basic idea is to decompose the solution to the CK Eq. (\ref{DL2}) according to
\begin{equation}
p_m(x,t)=p(x,t)\rho_m(x)+\epsilon w_n(x,t),
\end{equation}
where $\sum_{m=0,1}w_m(x,t)=0$. Using a Liapunov-Schmidt reduction one can derive the FP equation
\begin{equation}
\label{FP}
\frac{\partial p}{\partial t}=- \frac{\partial}{\partial x}(V(x)p)+\epsilon \frac{\partial}{\partial x}\left (D(x)\frac{\partial p}{\partial x}\right ),
\end{equation}
where we have dropped an $O(\epsilon)$ contribution to the drift term, and
\begin{align}
 D(x)
 &=\frac{4v^2\alpha(x)\beta(x)}{\alpha(x)+\beta(x)}.
 \label{Dxx}
\end{align}
Under this approximation, the position of the RTP evolves according to the stochastic differential equation
\begin{equation}
dX=V(X)dt+\sqrt{2\epsilon D(X)} dW(t),
\end{equation}
where $W(t)$ is a Wiener process with 
\begin{equation}
\langle W(t)\rangle =0,\quad \langle W(t)W(t')\rangle =\min\{t,t'\}.
\end{equation}
Given the specific form of the FP Eq. (\ref{FP}), the multiplicative noise is defined according to the kinetic interpretation of stochastic calculus.

Although the diffusion approximation is useful in capturing certain time-dependent aspects of the RTP, it breaks down in the large time limit. In particular, it yields a poor estimate of the stationary density of the exact model (\ref{DL2}). This point was originally highlighted within the context of molecular transport models \cite{Newby10a,Newby11}. The normalizability of the stationary density requires the corresponding flux to be zero for all $x\in \R$. In the case of the FP equation (\ref{FP}) this means
\[J(x)=-V(x)p(x)+\frac{\partial [D(x)p(x)]}{\partial x}=0,
\]
which yields the stationary density
\begin{equation}
p(x)={\mathcal N}\e^{-\Psi(x)/\epsilon},
\end{equation}
where
\begin{equation}
\Psi(x)=\int_{0}^x\frac{V(y)}{D(y)}dy.
\end{equation}
Clearly the quasipotential $\Psi(x)$ differs from the exact quasipotential $\Phi(x)$ of Eq. (\ref{sswkb}), resulting in exponentially significant errors for small $\epsilon$.
Following \cite{Newby11}, we can understand the source of this error by noting that the zero flux condition of the exact model (\ref{DL2}) implies 
\[J(x)=\sum_{n=0,1}v_np_n(x)=0,\]
that is, $p_0(x)=p_1(x)=p(x)/2$.
The underlying assumption of the QSS reduction is that the solution is close to the stationary distribution of the Markov chain, that is, $p_n(x)\sim \rho_n(x)$. Therefore, in order to be consistent with the exact zero flux condition, we would require $\sum_{n=0,1}v_n\rho_n(x)=V(x)=0$ for all $x$. This contradicts the fact that $V(x)$ only vanishes at $x=0$ for the switching rates (\ref{trap}) and at $x=\pm \bar{x}$ for the switching rates (\ref{trap1}). The problems with the diffusion approximation for an RTP also carry over to the calculation of the MFPT. 

Although the diffusion approximation breaks down in the long time limit, it can capture the behavior of a velocity jump process on shorter time-scales. For example, it would apply to FPT problems outside the weak noise regime where rare events dominate. This has been shown in a wide variety of models of motor-driven intracellular transport \cite{Bressloff13}. It is particularly useful when the number of velocity states are greater than two or transport occurs in more than one spatial dimension. Both of these latter features have been included in RTP models \cite{Mori19,Santra20,Basu20,Santra20a}. A more challenging problem is extending the asymptotic analysis of escape problems for RTPs with multiple internal states moving in two or more spatial dimensions. The first step would be to identify an appropriate mechanism for trapping.

\setcounter{equation}{0}
\renewcommand{\theequation}{A.\arabic{equation}}
\renewcommand{\thesubsection}{A.\arabic{subsection}}
\section*{Appendix A: Boundary layer analysis for the cubic potential.}

In this appendix we summarize the boundary layer analysis of Ref. \cite{Newby13a}, which leads to the result (\ref{c1}). Again the analysis simplifies greatly by focusing on the two-state RTP model rather than developing the theory for a general PDMP, which introduces additional technicalities. In order to deal with the blow up of the secular term in Eq. (\ref{An1}), we introduce an additional transition layer between the bulk or outer solution and the boundary layer. The scaling of this transition layer is determined by performing the change of variables $x=x_*-\epsilon^{\theta} y$, $0 < \theta < 1$, and defining
\begin{equation}
B_n(y)={\xi}_{0,n}(x_*-\epsilon^{\theta}y) .
\end{equation}
Introduce the asymptotic expansion
\begin{equation}
B_n(y)\sim B_n^{(0)}(y)+\epsilon^{s}B_n^{(1)}(y)+\epsilon^{2s} B_n^{(2)}(y),\quad s>0.
\end{equation}
Eq. (\ref{adj}) for $j=0$ becomes
\begin{align}
&\sum_{m=0,1}\bigg [\delta_{n,m}\epsilon^{1-\theta}v_n\frac{d}{d y}- \left [Q_{mn}(x_*)-\epsilon^{\theta}yQ'_{mn}(x_*)+\ldots \right ]\bigg ]\nonumber \\
&\quad \times \bigg (B_m^{(0)}(y)+\epsilon^{s}B_m^{(1)}(y)+\epsilon^{2s} B_m^{(2)}(y)\bigg)=0.
\label{adj2}
\end{align}
The $O(1)$ equation is
\begin{equation}
\sum_{m=0,1}Q_{mn}(x_*)B_m^{(0)}(y)=0,
\end{equation}
which implies that
\begin{equation}
B_n^{(0)}(y)=a_0(y)
\end{equation}
for some scalar function $a_0(y)$. The expansion (\ref{adj2}) then becomes
\begin{align}
&\epsilon^{1-\theta}v_na_0'(y)- \epsilon^{s}\sum_{m=0,1}Q_{mn}(x_*)B_m^{(1)}(y)+O(\epsilon)\nonumber \\
&\quad +o(\epsilon^{s},\epsilon^{1-\theta})=0.
\end{align}
This suggests taking $s=1-\theta$, which yields the $O(\epsilon^{1-\theta})$ equation
\begin{equation}
B_m^{(1)}(y)=-a_0'(y)\widehat{S}_m(x_*),
\end{equation}
where $\widehat{\bf S}(x_*)$ is the solution to Eq. (\ref{ztop}). Combining the results so far, we have
\begin{equation}
\label{Bn}
B_n(y)\sim a_0(y)-\epsilon^{1-\theta} a_0'(y)\widehat{S}_n(x_*).
\end{equation}

The next step is to calculate $a_0(y)$ by proceeding to higher order. We find 
\begin{align}
&-\epsilon^{2(1-\theta)} a_0''(y)v_n\widehat{S}_n(x_*)-\epsilon^{2(1-\theta)} \sum_{m=0,1}Q_{mn}(x_*)B_m^{(2)}(x_*)\nonumber \\
&\quad -\epsilon^{(1-\theta)\theta} ya_0'(y)\sum_{m=0,1}Q_{mn}'(x_*)\widehat{S}_m(x_*)=0.
\end{align}
Setting $\theta=1/2$ then yields
\begin{align}
&a_0''(y)v_n\widehat{S}_n(x_*)+ya_0'(y)\sum_{m=0,1}Q_{mn}'(x_*)\widehat{S}_m(x_*) \nonumber \\
&\quad  =-\sum_{m=0,1}Q_{mn}(x_*)B_m^{(2)}(x_*).
\end{align}
Multiplying both sides by $\rho_n(x_*)$, summing over $n$ and applying the Fredholm alternative theorem leads to the solvability condition
\begin{align}
\label{da0}
&a_0''(y)\sum_{n=0,1}\rho_n(x_*)v_n\widehat{S}_n(x_*)\nonumber \\
&\quad +ya_0'(y) \sum_{m,n=0,1}\rho_n(x_*)Q_{mn}'(x_*)\widehat{S}_m(x_*)=0.
\end{align}
In addition, ${\bf Q}(x_*)\rho(x_*)=0$ implies ${\bf Q}(x_*){\bm \rho}'(x_*)=-{\bf Q}'(x_*){\bm \rho}(x_*)$, and
\begin{align}
&\sum_{m,n=0,1} \widehat{S}_m(x_*)Q_{mn}(x_*)\rho_n'(x_*)\nonumber \\
&=\sum_{m,n=0,1} \widehat{S}_m(x_*)Q_{mn}(x_*)\rho_n'(x_*)=-\sum_{n=0,1}v_n\rho_n'(x_*).
\end{align}
Therefore, Eq. (\ref{da0}) reduces to the form
\begin{equation}
a_0''(y)+ya_0'(y)\frac{\sum_{n=0,1}v_n\rho_n'(x_*)}{\sum_{n=0,1}\rho_n(x_*)v_n\widehat{S}_n(x_*)}=0.
\end{equation}
Noting that $\sum_{n=0,1}v_n\rho_n'(x_*)=\overline{V}'(x_*)$ and
\[\sum_{n=0,1}\rho_n(x_*)v_n\widehat{S}_n(x_*)=\rho_0(x_*)v[\widehat{S}_0(x_*)-\widehat{S}_1(x_*)],\]
it follows that the fraction on the left-hand side is equal to $-\Phi''(x_*)$ so we have
\begin{equation}
a_0''(y)-ya_0'(y)\Phi''(x_*)=0.
\end{equation}
Exploiting the fact that $\Phi''(x_*)<0$, the solution for $a_0'(y)$ is
\[a_0'(y)=\bar{c}_1\e^{\Phi''(x_*)y^2/2}\]
and thus
\begin{equation}
a_0(y)=\bar{c}_0+\bar{c_1}\int_0^y \e^{\Phi''(x_*)u^2/2}du,
\end{equation}
where $\bar{c}_0,\bar{c}_1$ are integration constants.

Substituting the solution for $a_0(y)$ into (\ref{Bn}) and setting $\theta=1/2$ gives
\begin{align}
B_n(y)&\sim \bar{c}_0+\bar{c_1}\int_0^y \e^{\Phi''(x_*)u^2/2}du\nonumber \\
&\quad -\sqrt{\epsilon}\bar{c}_1\e^{\Phi''(x_*)y^2/2}\widehat{S}_n(x_*),
\end{align}
which replaces Eq. (\ref{An1}). This solution is bounded as $y\rightarrow \infty$ so it can be matched with the outer solution, that is,
$\lim_{y\rightarrow \infty}B_n(y)=1$. Hence,
\begin{equation}
\bar{c}_0+\bar{c}_1\int_0^{\infty} \e^{\Phi''(x_*)u^2/2}du=\bar{c}_0+\bar{c_1}\sqrt{\frac{\pi}{2|\Phi''(x_*)|}}=1.
\end{equation}
In addition, as $y\rightarrow 0$, we have
\begin{equation}
B_n(y)\sim \bar{c}_0+\bar{c_1}y-\sqrt{\epsilon}\bar{c}_1\widehat{S}_n(x_*).
\end{equation}
Matching with the boundary layer solution (\ref{An1}) then implies that $c_0=\bar{c}_0$ and $c_1=-\sqrt{\epsilon}\bar{c}_1$. Finally, if we impose the absorbing boundary condition $B_1(0)=0$ at $x=x_*$ and express $\bar{c}_0$ in terms of $\bar{c}_1$, then
\begin{align}
1-\bar{c_1}\sqrt{\frac{\pi}{2|\Phi''(x_*)|}}-\sqrt{\epsilon}\bar{c}_1\widehat{S}_n(x_*)=0.
\end{align}
On rearranging we recover Eq. (\ref{c1}).

\end{document}